\documentstyle[aps,prb,preprint]{revtex}

\begin{document}

\title{Pressure Calculation in Polar and Charged Systems using Ewald
  Summation: Results for the Extended Simple Point Charge Model of
  Water}

\author{Gerhard Hummer and Niels Gr{\o}nbech-Jensen}

\address{Theoretical Division, MS K710, Los Alamos National
  Laboratory, Los Alamos, New Mexico 87545, USA}

\author{Martin Neumann}

\address{Institut f\"{u}r Experimentalphysik, Universit\"{a}t Wien,
  A-1090 Vienna, Austria}

\date{LA-UR 98-1173; {\em J. Chem. Phys.}, in press, 15-AUG-1998}

\maketitle

\clearpage

\begin{abstract}
  Ewald summation and physically equivalent methods such as
  particle-mesh Ewald, kubic-harmonic expansions, or Lekner sums are
  commonly used to calculate long-range electrostatic interactions in
  computer simulations of polar and charged substances.  The
  calculation of pressures in such systems is investigated.  We find
  that the virial and thermodynamic pressures differ because of the
  explicit volume dependence of the effective, resummed Ewald
  potential.  The thermodynamic pressure, obtained from the volume
  derivative of the Helmholtz free energy, can be expressed easily for
  both ionic and rigid molecular systems.  For a system of rigid molecules,
  the electrostatic energy and the forces at the atom positions are
  required, both of which are readily available in molecular dynamics
  codes.  We then calculate the virial and thermodynamic pressures for
  the extended simple point charge (SPC/E) water model at standard
  conditions.  We find that the thermodynamic pressure exhibits
  considerably less system size dependence than the virial pressure.
  From an analysis of the cross correlation between the virial and
  thermodynamic pressure, we conclude that the thermodynamic pressure
  should be used to drive volume fluctuations in constant-pressure
  simulations.
\end{abstract}

\clearpage

\section{Introduction}

Pressure is one of the fundamental thermodynamic variables.  The
calculation of pressures in fluid systems using computer simulations
is generally considered to be a routine task.  However, difficulties
arise in the presence of long-range interactions.  Here, we
investigate the calculation of pressures in computer simulations of
charged and polar systems, where the long-range Coulomb interactions
are commonly treated with Ewald lattice summation\cite{Ewald:21} or
physically equivalent methods like particle-mesh Ewald,\cite{Darden:93}
kubic-harmonic expansions,\cite{Slattery:80} or Lekner
sums.\cite{Lekner:91,Jensen:96,Jensen:97:2} A mechanistic definition
of the pressure leads to the standard virial expression.  A
thermodynamic definition of the pressure is based on the volume
dependence of the Helmholtz free energy.  When the Coulomb
interactions are resummed by using, e.g., the Ewald method, the
resulting effective pair interactions depend explicitly on the volume.
In addition, self interactions are present that also depend on the
volume.  As a consequence, the virial and thermodynamic pressures are
not identical for finite Coulomb systems, even though the two
pressures are expected to converge in the thermodynamic limit.

The paper is organized as follows: in section \ref{sec:theory}, we
derive expressions for the virial and thermodynamic pressures.  In
section \ref{sec:pressCoul}, we study the pressure in systems with
long-range Coulomb interactions. For the thermodynamic pressure, we
derive a simple formula that can be readily implemented in standard
molecular dynamics or Monte Carlo codes.  In sections \ref{sec:sim}
and \ref{sec:spce}, we study the system size dependence of the virial
and thermodynamic pressures for the extended simple point charge (SPC/E) water
model\cite{Berendsen:87} under standard conditions.

\section{Virial and thermodynamic pressures}
\label{sec:theory}
\subsection{Virial pressure}
The pressure $p$ can be calculated from a mechanistic prescription
equating the exterior and interior forces on the container.  This
leads to the virial expression for the
pressure in an atomic system,\cite{Goldstein:80:virial}
\begin{eqnarray}
  \label{eq:virial}
  p_V & = & \rho k_{\rm B} T + \frac{1}{3V} \left\langle \sum_i {\bf
 r}_i \cdot {\bf F}_i \right\rangle~,
\end{eqnarray}
where $\rho$ is the number density of particles; $k_{\rm B}$ is
Boltzmann's constant; $T$ is the temperature ($k_{\rm B} T =
\beta^{-1}$); and $V$ is the volume.  The sum extends over the scalar
product betwen particle positions ${\bf r}_i$ and forces ${\bf F}_i$
exerted on particle $i$ due to other particles in the system.
$\langle \ldots \rangle$ denotes a canonical average.  For computer
simulations under periodic boundary conditions with pair forces, it is
convenient to rewrite the virial equation in a form that makes the
translational invariance explicit:
\begin{eqnarray}
  \label{eq:virial_pair}
  p_V & = & \rho k_{\rm B} T - \frac{1}{3V} \left\langle
  \sum_{i,j \atop i<j} {\bf F}_{ij} \cdot {\bf r}_{ij} \right\rangle~,
\end{eqnarray}
where ${\bf r}_{ij}={\bf r}_j - {\bf r}_i$ and ${\bf F}_{ij}=-\partial
v({\bf r}_{ij})/\partial {\bf r}_i = \partial v({\bf r}_{ij})/\partial
{\bf r}_{ij}$ is the pair force exerted on particle $i$ by particle
$j$, derived from a pair potential $v({\bf r})$; and the sum is over
all pairs of particles in the system.

For a system of rigid polyatomic molecules $i, j$ with interaction
sites $\alpha$ and $\beta$, one obtains an analogous formula when
the forces ${\bf F}_{i\alpha j\beta}$ between molecular sites
are projected onto a vector ${\bf r}_{ij}$ between the ``centers'' of
the two molecules (e.g., the center of mass),
\begin{eqnarray}
  \label{eq:virial_pair_mol}
  p_V & = & \rho k_{\rm B} T - \frac{1}{3V} \left\langle
    \sum_{i,j \atop i<j} \sum_{\alpha,\beta} {\bf F}_{i\alpha j\beta}
    \cdot {\bf r}_{ij} \right\rangle\nonumber\\
  & = & \rho k_{\rm B} T - \frac{1}{3V} \left\langle
    \sum_{i,j \atop i<j} {\bf F}_{ij}
    \cdot {\bf r}_{ij} \right\rangle~.
\end{eqnarray}
Here, ${\bf F}_{ij}$ is the net force between two rigid molecules, summed
over molecular sites $\alpha$ and $\beta$.  Otherwise, the constraint
forces maintaining the rigidity of the molecules have to be included
explicitly in Eq.~(\ref{eq:virial_pair_mol}).

\subsection{Thermodynamic pressure}

The thermodynamic expression for the pressure is derived from the
relation between the pressure $p_T$, the Helmholtz free energy $F$, and
the volume $V$,
\begin{eqnarray}
  \label{eq:p_therm}
  p_T & = & -\left(\frac{\partial F}{\partial V}\right)_{T,N}~.
\end{eqnarray}
The statistical-mechanical relation between the free energy $F$ and
the partition function $Q_N(V,T)$ for $N$ identical classical particles in a
canonical ensemble is:
\begin{eqnarray}
  \label{eq:part}
Q_N(V,T) & = & e^{-\beta F} = \frac{1}{h^{3N} N!}\int\int
e^{-\beta H({\bf r}^N,{\bf p}^N)} d{\bf r}^N d{\bf p}^N~,
\end{eqnarray}
where $h$ is Planck's constant; $H=K+U$ is the Hamiltonian; and the
integration extends over the positions ${\bf r}^N$ and momenta ${\bf
  p}^N$ of all particles.  In taking the volume derivative
Eq.~(\ref{eq:p_therm}), the kinetic energy $K$ is independent of the
volume.  Transforming the positional coordinates into dimensionless
form, $V^{-N} {\bf r}^N$, and pulling out a factor $V^N$ from the
integral, leads to the ideal gas term $\rho k_{\rm
  B}T$ for the pressure.  The non-ideal contributions are contained in
the volume dependence of the potential energy $U$,
\begin{eqnarray}
  \label{eq:pcanon}
  p_T & = & \rho k_{\rm B}T - \left\langle \frac{\partial U}{\partial
  V}\right\rangle~.
\end{eqnarray}
Typically, $U$ does not depend explicitly on the volume.  The volume
dependence of $U$ then arises from the volume scaling of the particle
positions.  In the absence of an explicit volume dependence, we can
express $\partial U/\partial V$ as
\begin{eqnarray}
  \label{eq:dudv}
  \frac{\partial U}{\partial V} & = & \sum_{i} \frac{\partial U}{\partial {\bf
      r}_i} \cdot \frac{\partial {\bf r}_i}{\partial V} = \frac{1}{3V}
      \sum_i \frac{\partial U}{\partial {\bf r}_i} \cdot {\bf r}_i~,
\end{eqnarray}
with $\partial {\bf r}_i/\partial V ={\bf r}_i/3V$.  By using ${\bf
  F}_i = - \partial U/\partial {\bf r}_i$ and combining
Eqs.~(\ref{eq:pcanon}) and (\ref{eq:dudv}), we find the corresponding
thermodynamic pressure to be equivalent to the mechanistic pressure
Eq.~(\ref{eq:virial}).

\section{Pressure in systems with long-range Coulomb interactions}
\label{sec:pressCoul}

\subsection{Thermodynamic pressure in ionic systems}

The identity between the virial and thermodynamic pressures,
Eqs.~(\ref{eq:virial}) and (\ref{eq:pcanon}), does not hold if the
potential depends explicitly on the system volume.  Such an explicit
volume dependence arises in computer simulations of charged and polar
substances, when the long-range Coulomb interactions are resummed,
e.g., by using the Ewald method.\cite{Ewald:21}

We first split the total potential energy $U$ into a short-range part
$U^{\rm (sr)}$ and a long-range Coulomb part $U^{\rm (c)}$,
\begin{eqnarray}
  \label{eq:Usplit}
  U & = & U^{\rm (sr)} + U^{\rm (c)}~.
\end{eqnarray}
The pressure $p^{\rm (sr)}$ corresponding to $U^{\rm (sr)}$ contains
the ideal-gas term and the contributions from short-range pair
interactions,
\begin{eqnarray}
  \label{eq:pshort}
  p^{\rm (sr)} & = & \rho k_{\rm B} T -\frac{1}{3V} \sum_{i,j \atop
    i<j}\sum_{\alpha,\beta} {\bf F}_{i\alpha j\beta}^{\rm (sr)}
  \cdot {\bf r}_{ij}~,
\end{eqnarray}
where the short-range forces ${\bf F}_{i\alpha j\beta}^{\rm (sr)}$ are
those derived from the short-range part $U^{\rm (sr)}$ of the
potential energy.  Note that the virial and thermodynamic expressions
for $p^{\rm (sr)}$ are equivalent, and therefore the subscripts ``{\em
  V}'' or ``{\em T}'' have been omitted in Eq.~(\ref{eq:pshort}).

Next, we consider the pressure arising from the potential energy
$U^{\rm (c)}$ of long-range Coulomb interactions.  In Ewald lattice
summation, the charges in a periodically replicated simulation box
interact with an effective potential.  That potential is obtained from
a summation over all periodic images.  In addition, a self interaction
arises from interactions with a particle's own images.  This leads to
a Coulomb energy $U^{\rm (c)}$ for a system of partial charges
$q_{i\alpha}$ at positions ${\bf r}_{i\alpha}$:
\begin{eqnarray}
  \label{eq:Uew}
  U^{\rm (c)} & = & \sum_{i,j \atop i<j}\sum_{\alpha,\beta}
  q_{i\alpha} q_{j\beta}
  \varphi({\bf r}_{i\alpha j\beta})\nonumber\\&& + \sum_{i}
  \sum_{\alpha,\beta \atop \alpha<\beta} q_{i\alpha} q_{i\beta}
  \left[\varphi({\bf r}_{i\alpha
  i\beta})-\frac{1}{|{\bf r}_{i\alpha i\beta}|}\right]\nonumber\\&&
 + \frac{1}{2} \sum_{i}
  \sum_{\alpha} {q_{i\alpha}}^2 \lim_{{\bf r}\rightarrow 0}
  \left[\varphi({\bf r})-\frac{1}{|{\bf r}|}\right]~.
\end{eqnarray}
The first sum is the intermolecular contribution; the second and
third sums are the intramolecular contributions, with the self
interactions contained in the third sum.  $\varphi({\bf r})$ is the
effective, resummed Coulomb potential, with a Fourier
representation:\cite{Brush:66}
\begin{eqnarray}
  \label{eq:phi_Four}
  \varphi({\bf r}) = \frac{1}{V}\sum_{{\bf k} \atop k\neq
  0}\frac{4\pi}{{\bf k}^2} e^{i {\bf k}\cdot{\bf r}}~,
\end{eqnarray}
where the ${\bf k}$ sum extends over the reciprocal lattice
corresponding to the lattice vectors ${\bf n}$ of periodically
replicated simulation boxes.  In a cubic lattice of length
$L=V^{1/3}$, we have ${\bf n}=L\,(i,j,k)$, and ${\bf k}=2\pi
L^{-1}\,(i,j,k)$, where $i$, $j$, and $k$ are integers.  It is
numerically convenient to partly transform $\varphi({\bf r})$ into
real space, leading to its Ewald lattice sum representation,
\begin{eqnarray}
  \label{eq:phi}
  \varphi({\bf r}) & = & \sum_{\bf n}\frac{{\rm erfc}(\eta|{\bf r} + 
    {\bf n}|)} {|{\bf r}+{\bf n}|} + \sum_{{\bf k} \atop k\neq 0}
\frac{4\pi}{V k^2} e^{ -k^2/4\eta^2 + i{\bf k}\cdot{\bf r}}
-\frac{\pi}{V \eta^2}~.
\end{eqnarray}
$\eta$ is a convergence parameter chosen to accelerate numerical
convergence.  The value of $\varphi({\bf r})$ is independent of
$\eta$,\cite{Hummer:95:CPL}
\begin{eqnarray}
  \label{eq:dphideta}
  \frac{\partial \varphi({\bf r})}{\partial\eta} \equiv 0~.
\end{eqnarray}
The self-interactions
in $U^{\rm (c)}$ are given by the interactions of a unit point charge with its
periodic images, subtracting the bare self interaction, $\varphi({\bf
  r})-1/|{\bf r}|$, with the appropriate limit taken for ${\bf
  r}\rightarrow 0$.  For a given box shape, $\varphi({\bf r})$ scales
with the box volume $V$ as
\begin{mathletters}
  \label{eq:phiscale}
  \begin{eqnarray}
    \varphi({\bf r}) & =  & V^{-1/3} \varphi^*({\bf r}^*)~,\\
    \frac{\partial\varphi({\bf r})}{\partial V} & = & -\frac{1}{3V}
    \varphi({\bf r})~,
  \end{eqnarray}
\end{mathletters}
where star superscripts denote volume-independent quantities.  This
follows from Eq.~(\ref{eq:phi_Four}) with volume scaling ${\bf r}\sim
V^{1/3}$ and ${\bf k}\sim V^{-1/3}$.  The same scaling is true
trivially for the direct $1/|{\bf r}|$ interactions.
For an ionic system of point
charges without bond constraints, Eqs.~(\ref{eq:pcanon}) and
(\ref{eq:phiscale}) immediately lead to an expression for the
thermodynamic pressure in terms of the Coulomb energy $U^{\rm (c)}$,
\begin{eqnarray}
  \label{eq:pionic}
  p_T & = & p^{\rm (sr)} + \frac{\langle U^{\rm (c)}\rangle}{3V}~.
\end{eqnarray}
Equation~(\ref{eq:pionic}) gives the well-known relation between the
pressure and energy of an ionic system, for which the Coulomb energy
is a homogeneous function of degree $-1$ in the
coordinates.\cite{Landau:80:Coulomb}

\subsection{Thermodynamic pressure in systems of rigid polyatomic molecules}

For a system of rigid molecules, we find the following volume scaling:
\begin{mathletters}
  \label{eq:molscal}
  \begin{eqnarray}
    \varphi({\bf r}_{i\alpha j\beta}) & = & V^{-1/3}\varphi^*({\bf r}_{ij}^*
    + V^{-1/3} {\bf d}_{i\alpha j\beta})~,\\
    \frac{\partial\varphi({\bf r}_{i\alpha j\beta})}{\partial V} & = &
    -\frac{1}{3V} \left[ \varphi({\bf r}) + \frac{\partial\varphi({\bf
          r}_{i\alpha j\beta})}{\partial {\bf r}_{i\alpha j\beta}}\cdot {\bf
        d}_{i\alpha j\beta} \right]~,\\
    \frac{\partial}{\partial V} \frac{1}{|{\bf r}_{i\alpha i\beta}|} &
    = & -\frac{1}{3V}\left[\frac{1}{|{\bf r}_{i\alpha i\beta}|} +
    \frac{\partial}{\partial {\bf r}_{i\alpha i\beta}}
    \frac{1}{|{\bf r}_{i\alpha i\beta}|}\cdot {\bf
        d}_{i\alpha i\beta}\right] \equiv 0~,
    \label{eq:drv}
  \end{eqnarray}
\end{mathletters}
where ${\bf r}_{ij}$ is the distance vector between two molecule
centers; ${\bf d}_{i\alpha}={\bf r}_{i\alpha}-{\bf r}_i$ is the vector
from the center to site $\alpha$; and ${\bf d}_{i\alpha j\beta}={\bf
  d}_{j\beta} - {\bf d}_{i\alpha}$.  Equation~(\ref{eq:drv}) follows from
the volume independence of the intramolecular distance vector ${\bf
  r}_{i\alpha i\beta} = {\bf d}_{i\alpha i\beta}$.  Combining
Eqs.~(\ref{eq:Uew}) and (\ref{eq:molscal}), we find
for the volume derivative of the Coulomb energy:
\begin{eqnarray}
  \label{eq:dudvEW}
  \frac{\partial U^{\rm (c)}}{\partial V} & = & -\frac{U^{\rm
    (c)}}{3V} -\frac{1}{3V}\left\{
    \sum_{i,j \atop i<j}\sum_{\alpha,\beta}
    \left[\frac{\partial}{\partial {\bf r}_{i\alpha j\beta}}
      q_{i\alpha} q_{j\beta} \varphi({\bf r}_{i\alpha j\beta})\right]
    \cdot {\bf d}_{i\alpha j\beta}\right.
  \nonumber\\&&\left. + \sum_{i}
    \sum_{\alpha,\beta \atop \alpha<\beta}
    \left[\frac{\partial}{\partial {\bf r}_{i\alpha i\beta}}
      q_{i\alpha} q_{i\beta}  \left(\varphi({\bf r}_{i\alpha
          i\beta})-\frac{1}{|{\bf r}_{i\alpha i\beta}|}\right)\right]
    \cdot {\bf d}_{i\alpha i\beta}
  \right\}~.
\end{eqnarray}
We can simplify $\partial U^{\rm (c)}/\partial V$ further by expressing it in
terms of the intermolecular forces ${\bf F}^{\rm (inter)}_{i\alpha
  j\beta}$ exerted by site ${j\beta}$ onto site ${i\alpha}$,
\begin{eqnarray}
  \label{eq:forces_inter}
  {\bf F}^{\rm (inter)}_{i\alpha j\beta} & = & \frac{\partial}
  {\partial {\bf r}_{i\alpha j\beta}}   q_{i\alpha} q_{j\beta}
  \varphi({\bf r}_{i\alpha j\beta}) = - {\bf F}^{\rm (inter)}_{j\beta
  i\alpha}~,
\end{eqnarray}
and the intramolecular forces ${\bf F}^{\rm (intra)}_{i\alpha i\beta}$,
\begin{eqnarray}
  \label{eq:forces_intra}
  {\bf F}^{\rm (intra)}_{i\alpha i\beta} & = & \frac{\partial}
  {\partial {\bf r}_{i\alpha i\beta}} q_{i\alpha} q_{i\beta} \left[
    \varphi({\bf r}_{i\alpha i\beta})-\frac{1}{|{\bf r}_{i\alpha
        i\beta}|}\right]  = - {\bf F}^{\rm (intra)}_{i\beta
    i\alpha}~.
\end{eqnarray}
This leads to
\begin{eqnarray}
  \frac{\partial U^{\rm (c)}}{\partial V} & = & -\frac{1}{3V}
    \left(U^{\rm (c)} +
    \sum_{i,j \atop i<j} \sum_{\alpha,\beta} {\bf
      F}^{\rm (inter)}_{i\alpha j\beta} \cdot {\bf d}_{i\alpha j\beta}  +
    \sum_{i} \sum_{\alpha,\beta \atop \alpha<\beta}
      {\bf F}^{\rm (intra)}_{i\alpha i\beta}
      \cdot {\bf d}_{i\alpha i\beta}\right)~,
\end{eqnarray}
The sums over pairs of sites $i\alpha$ and $j\beta$ can be rewritten
as a single sum over all sites.  This is possible because the
distances ${\bf d}_{i\alpha j\beta} = {\bf d}_{j\beta} - {\bf
  d}_{i\alpha}$ are intramolecular and are continuous when a particle
crosses the box boundary (i.e., ${\bf d}_{i\alpha j\beta}$ does not
change when the periodic images of the particles $i$ or $j$ are used).
This results in
\begin{eqnarray}
  \frac{\partial U^{\rm (c)}}{\partial V} & = & -\frac{1}{3V} \left[
    U^{\rm (c)} -
    \sum_{i} \sum_{\alpha} \left( {\bf F}^{\rm (inter)}_{i\alpha}
    + {\bf F}^{\rm (intra)}_{i\alpha}\right) \cdot {\bf d}_{i\alpha}\right]~,
\end{eqnarray}
where ${\bf F}^{\rm (inter)}_{i\alpha}$ is the net intermolecular force on
site $i\alpha$,
\begin{eqnarray}
  \label{eq:force}
  {\bf F}^{\rm (inter)}_{i\alpha} & = & \sum_{j \atop j\neq i}
  \sum_\beta {\bf F}^{\rm (inter)}_{i\alpha j\beta}~,
\end{eqnarray}
and ${\bf F}^{\rm (intra)}_{i\alpha}$ is the net intramolecular force on
site $i\alpha$,
\begin{eqnarray}
  {\bf F}^{\rm (intra)}_{i\alpha} & = & \sum_{\beta \atop \beta\neq \alpha}
  {\bf F}^{\rm (intra)}_{i\alpha i\beta}~.
\end{eqnarray}
It is advantageous to add the intra and intermolecular forces because
in common Ewald-sum implementations the Fourier term already contains
the sum of both inter and intramolecular contributions which are thus
not easily separated.  We define the net Coulomb force ${\bf F}^{\rm
  (c)}_{i\alpha}$ on site $i\alpha$ as the sum of the inter and
intramolecular forces,
\begin{eqnarray}
  \label{eq:frc}
  {\bf F}^{\rm (c)}_{i\alpha} & = & {\bf F}^{\rm (inter)}_{i\alpha} +
  {\bf F}^{\rm 
  (intra)}_{i\alpha} = -\frac{\partial U^{\rm (c)}}{\partial {\bf
  r}_{i\alpha}}~.
\end{eqnarray}
We then find for the thermodynamic pressure of a system of rigid
molecules:
\begin{eqnarray}
  \label{eq:pmol}
  p_T & = & p^{\rm (sr)} + \frac{1}{3V} \left( \langle U^{\rm (c)}\rangle -
  \left\langle \sum_i \sum_\alpha {\bf F}^{\rm ({\rm c})}_{i\alpha}
  \cdot {\bf d}_{i\alpha} \right\rangle\right)~.
\end{eqnarray}
Thus the presence of intramolecular constraints in rigid polyatomic
molecules resulted in a force term to be subtracted from the pressure
of the purely ionic system, Eq.~(\ref{eq:pionic}).  Note that the
forces ${\bf F}^{\rm (c)}_{i\alpha}$ in Eq.~(\ref{eq:pmol}) are
derived from the Coulomb energy $U^{\rm (c)}$ alone.  Additional
ideal-gas and short-range contributions to the pressure are reflected
in $p^{\rm (sr)}$.

\subsection{Tin-foil boundary conditions and reaction field correction}

The infinite Ewald lattice is implicitly embedded in a conducting
medium with dielectric constant $\epsilon_{\rm rf} = \infty$,
corresponding to ``tin-foil'' boundary conditions.  This is the
appropriate choice for a conducting medium.  However, for a polar
substance it can be advantageous to use a reaction-field dielectric
constant $\epsilon_{\rm rf}$ similar to that of the bulk
medium.\cite{Boresch:97} The resulting correction to the Coulomb
energy $U^{\rm (c)}(\epsilon_{\rm rf}=\infty)$ is\cite{deLeeuw:80:a}
\begin{eqnarray}
  \label{eq:eps_rf}
  U^{\rm (rf)} & = & U^{\rm (c)}(\epsilon_{\rm rf}) - U^{\rm
  (c)}(\epsilon_{\rm rf}=\infty) = 
  \frac{2\pi}{(2\epsilon_{\rm rf}+1)V} {\bf M}^2~.
\end{eqnarray}
${\bf M}$ is the instantaneous dipole moment of the simulation volume
arising from the dipole moments ${\bf m}_i$ of individual molecules,
\begin{eqnarray}
  \label{eq:M}
  {\bf M} & = & \sum_i {\bf m}_i = \sum_i \sum_\alpha
  q_{i\alpha}{\bf d}_{i\alpha}~.
\end{eqnarray}
For rigid molecules the ${\bf m}_i$ do not change with volume.  The
reaction-field correction, Eq.~(\ref{eq:eps_rf}), thus scales as
$V^{-1}$,
\begin{eqnarray}
  \label{eq:dumdv}
  \frac{\partial U^{\rm (rf)}}{\partial V} & = & -\frac{U^{\rm (rf)}}{V}~.
\end{eqnarray}
The forces derived from the reaction-field correction are
\begin{eqnarray}
  \label{eq:dumdr}
  {\bf F}^{\rm (rf)}_{i\alpha} & = & - \frac{\partial U^{\rm
  (rf)}}{\partial{\bf r}_{i\alpha}} =
  - \frac{4\pi q_{i\alpha}}{(2\epsilon_{\rm rf}+1)V} {\bf M}~.
\end{eqnarray}
By using Eq.~(\ref{eq:M}), we can express the sum of reaction-field
forces projected onto the intramolecular distance vectors in terms of
the reaction-field energy $U^{(\rm rf)}$,
\begin{eqnarray}
  \label{eq:sumrf}
  \sum_{i,\alpha} {\bf F}^{\rm (rf)}_{i\alpha}\cdot {\bf
  d}_{i\alpha} &=& - 2 U^{\rm (rf)}~.
\end{eqnarray}
Accordingly, the volume derivative of the reaction-field energy
$U^{\rm (rf)}$ can be written as
\begin{eqnarray}
  \label{eq:durfdv}
  -\frac{\partial U^{\rm (rf)}}{\partial V} & = & \frac{1}{3V}
  \left[U^{\rm (rf)}-\sum_{i,\alpha} {\bf F}^{\rm
  (rf)}_{i\alpha}\cdot {\bf d}_{i\alpha}\right]~.
\end{eqnarray}
The correction Eq.~(\ref{eq:eps_rf}) for a finite reaction-field
dielectric constant $\epsilon_{\rm rf}$ then leads to an expression
for the thermodynamic pressure analogous to Eq.~(\ref{eq:pmol}),
\begin{eqnarray}
  \label{eq:pmol_M}
  p_T & = & p^{\rm (sr)} + \frac{1}{3V}\left[\langle U^{\rm
    (c)}(\epsilon_{\rm rf})
    \rangle - \left\langle \sum_i \sum_\alpha {\bf
      F}^{{\rm (c,}\epsilon_{\rm rf})}_{i\alpha} \cdot {\bf
    d}_{i\alpha} \right\rangle\right]~.
\end{eqnarray}
Here, the forces ${\bf F}^{{\rm (c,}\epsilon_{\rm rf})}_{i\alpha}$ are
derived from the Coulomb energy $U^{\rm (c)}(\epsilon_{\rm rf})$
\begin{eqnarray}
  \label{eq:fcrf}
  {\bf F}^{{\rm (c,}\epsilon_{\rm rf})}_{i\alpha} & = & -
  \frac{\partial U^{\rm (c)}(\epsilon_{\rm rf})}{\partial {\bf
  r}_{i\alpha}}~,
\end{eqnarray}
and contain the reaction field contribution ${\bf F}^{\rm (rf)}_{i\alpha}$
defined in Eq.~(\ref{eq:dumdr}).

\subsection{General considerations}

We emphasize the simplicity of the pressure expressions
Eqs.~(\ref{eq:pmol}) and (\ref{eq:pmol_M}) for systems of rigid
molecules.  The Coulomb energy contribution $\langle U^{\rm
  (c)}\rangle/3V$ is analogous to that of the corresponding ionic
system, Eq.~(\ref{eq:pionic}), corrected for the presence of
constraint forces.  A more or less equivalent expression for the
pressure in Coulombic systems treated with Ewald summation was derived
before by Smith,\cite{Smith:87} and similarly by Boulougouris {\em et
al.}\cite{Boulougouris:98}.  However, in those derivations the volume
derivative was carried out explicitly for the Ewald energy.  Also, the
derivations start from an approximate expression for the Ewald energy
that does not include the full real-space lattice sum and self terms.
Therefore, the derivations did not arrive at a closed expression and
the simplicity of the results given here was masked.

Expressions for the pressure tensor ${\bf P}$ for Ewald summation were
derived previously by Nos\'{e} and Klein,\cite{Nose:83} and
Heyes,\cite{Heyes:94:PRB} as discussed by Alejandre {\em et
al.},\cite{Alejandre:95} as well as by
others.\cite{Brown:95,Taylor:97,Smith:94,Essmann:95} However, the
tensor character does not lend itself easily to a compact notation for
the bulk pressure $p = {\rm Tr}({\bf P})$ in a homogeneous system.
Equations~(\ref{eq:pionic}), (\ref{eq:pmol}), and (\ref{eq:pmol_M})
have the advantage of being independent of the specific method used to
evaluate the energies and forces.  All that is needed is the total
Coulomb energy and forces at all sites that are consistent with that
energy.  This is what molecular dynamics codes will normally produce
at no additional cost.  The Coulomb interactions can then be evaluated
by using conventional Ewald sums,\cite{Ewald:21,Allen:87}
particle-mesh Ewald,\cite{Darden:93} kubic-harmonic
expansion\cite{Slattery:80}, or Lekner
sums.\cite{Lekner:91,Jensen:96,Jensen:97:2} For approximate Coulomb
energy calculations such as reaction-field\cite{Barker:73,Hummer:92:b}
or generalized reaction-field methods,\cite{Hummer:94:e}
Eqs.~(\ref{eq:pionic}), (\ref{eq:pmol}), and (\ref{eq:pmol_M}) suggest
an evaluation of the pressure that is formally consistent with that of
Ewald sums and physically equivalent methods.

\section{Computer simulations}
\label{sec:sim}
To investigate the quantitative differences between the virial and
thermodynamic pressures, we study a model of water at standard
conditions (298~K temperature, 997.07 kg m$^{-3}$ mass density
corresponding to a number density of $\rho=33.33$ nm$^{-3}$).  We
use the SPC/E model of water,\cite{Berendsen:87} formed by a
Lennard-Jones center on the oxygens,
\begin{eqnarray}
  \label{eq:vOO}
  v_{\rm LJ}(r) & = & \frac{A}{r^{12}}-\frac{B}{r^6}~,
\end{eqnarray}
where $A=0.3428^{12}$ kJ nm$^{12}$/mol and $B=0.37122^6$ kJ
nm$^6$/mol.  In addition, the SPC/E model carries three partial
charges.  The hydrogen and oxygen sites carry charges $q_H=0.4238 e$
and $q_O=-2q_H$, respectively, where $e$ is the elementary charge.
The oxygen-hydrogen bond length is 0.1 nm, the
hydrogen-oxygen-hydrogen bond angle is $\cos^{-1}(-1/3)\approx 109.47$
deg.

We use Metropolis Monte Carlo simulations for the canonical
sampling,\cite{Allen:87} where the translational and rotational move
widths are chosen to give an acceptance rate of about 40 per cent.
Ewald summation is used for the electrostatic interactions, with
$\eta=5.6/L$ where $L=V^{1/3}$ is the length of the cubic box.  A
spherical cutoff of $L/2$ is used for the real space interactions
(charge and Lennard-Jones).  The real-space potentials are shifted by
a constant, such that they are zero at the cutoff.  The Fourier space
sum is truncated at ${\bf k}^2 \leq 38 (2\pi/L)^2$, leading to
$2\times 510$ ${\bf k}$ vectors being considered.  A reaction-field
dielectric constant of $\epsilon_{\rm rf} = 65$ has been used in all
simulations.  Standard finite-size corrections were applied to the
Lennard-Jones contributions to pressure and potential
energy.\cite{Allen:87}

System sizes of $N=16$, 32, 64, 128, 256, and 512 water molecules are
studied.  Starting from random configurations, these systems have been
equilibrated for at least 250\,000 Monte Carlo passes.  (One pass
corresponds to one attempted move for each of the $N$ particles.)  In
the production runs, the energy as well as the virial and
thermodynamic pressures are calculated every tenth pass.

\section{Results for SPC/E water}
\label{sec:spce}
Table~\ref{tab:p} contains the simulation characteristics, as well as
results for the virial and thermodynamic pressures.  The thermodynamic
pressure is calculated using Eq.~(\ref{eq:pmol_M}).  The virial
pressure is calculated using Eq.~(\ref{eq:virial_pair_mol}), where the
pair forces are derived from the total potential energy $U=U^{\rm
  (sr)}+U^{\rm (c)}$.  Also included in Table~\ref{tab:p} are results
for the average potential energy per particle.  (To compare with the
experimental energy, one has to add a polarization correction of 5.22
kJ/mol.\cite{Berendsen:87}) Errors are obtained from a block
analysis,\cite{Allen:87} plotting calculated standard deviations of
the mean as a function of the number of blocks used.  The reported
error is then the plateau value reached in the limit of long blocks
with typically more than about 10\,000 Monte Carlo passes.

The system size dependence of the virial and thermodynamic pressure is
shown in Fig.~\ref{fig:p}.  From Table~\ref{tab:p} and
Fig.~\ref{fig:p}, we find that $p_V$ and $p_T$ converge to the same
value of about $-5$ MPa (1 MPa = 10 bar) for large system sizes, with
a statistical error of about 2 MPa.  This convergence is expected as
the thermodynamic and virial pressure should be identical in the
thermodynamic limit.  However, the thermodynamic pressure exhibits a
considerably weaker system size dependence than the virial pressure.
The thermodynamic pressure for as few as 64 SPC/E water molecules is
in agreement with large system sizes.  The virial pressure, on the
other hand, scales as roughly $1/N^2$ for small to intermediate system
sizes, with its value off by about one order of magnitude for $N=64$.
We emphasize that for typical system sizes of $N\geq 256$, the virial
and thermodynamic pressures are identical within statistical errors
for SPC/E water under standard conditions.

Figure~\ref{fig:gXX} shows the radial distribution functions of water
oxygens and hydrogens, which were calculated also in the corners of
the cubic simulation box with appropriate weights.  We find that the
the radial distribution functions for $N\geq 64$ water molecules are
practically indistinguishable, whereas the $N=16$ and $N=32$
simulations are somewhat more structured beyond the first peaks.
These slight structural differences could explain the deviations of
the thermodynamic pressure for those small system sizes.  We caution
that these are results for the specific thermodynamic state (room
temperature and standard density) studied here, and we expect more
pronounced finite-size effects, e.g., for low densities.

In constant pressure simulations,\cite{Parrinello:80,Andersen:80} the
box volume is rescaled according to the ``instantaneous pressure''
obtained from individual configurations by omitting the canonical
average $\langle\ldots\rangle$ in the pressure formulas above.  It is
therefore important that not only the average but also the
instantaneous pressure driving the volume fluctuations be correct.  As
measures of discrepancy between the virial and thermodynamic
pressures $p_V$ and $p_T$, we use the correlation coefficient $r$ and
the average absolute deviation minus the deviation of the averages,
$\Delta$,
\begin{mathletters}
  \label{eq:deviation}
  \begin{eqnarray}
    r & = & \frac{\left\langle(p_T-\langle p_T\rangle)(p_V-\langle
        p_V\rangle)\right\rangle}
    {\left\langle (p_T-\langle p_T\rangle)^2\right\rangle^{1/2}
      \left\langle (p_V-\langle p_V\rangle)^2\right\rangle^{1/2}}~,\\
    \Delta & = & \left\langle | p_T-\langle p_T\rangle - p_V+\langle
    p_V\rangle | \right\rangle~,
  \end{eqnarray}
\end{mathletters}
where instantaneous pressures $p_T$ and $p_V$ are used.  Results for
$r$ and $\Delta$ are listed in Table~\ref{tab:p}.  The
cross-correlation coefficient $r$ indicates strong correlation, with
$r$ values between 0.90 and 0.997 for $N=16$ to $N=512$.  However, the
average absolute deviation $\Delta$ between the two pressures is
significant even for systems of 512 water molecules, scaling
approximately as $\Delta\sim 1/N$.  Therefore, in constant pressure
simulations, the use of the thermodynamic pressure appears
advantageous.

In an earlier study of pressure effects on the stability of
hydrophobic aggregates in water,\cite{Hummer:PNAS:98} we determined
the thermodynamic pressure of SPC water\cite{Berendsen:81} as a
function of density using Eq.~(\ref{eq:pmol_M}).  For the temperature
and density studied here ($T=298$ K, $\rho=33.33$ nm$^{-3}$), we found
a pressure of about $37\pm 6$ MPa for SPC water.  From the density
dependence of the pressure, we determined a compressibility factor
$\rho k_{\rm B}T \chi_T\approx 0.06$ for SPC water, where $\chi_T$ is
the isothermal compressibility.  That compressibility factor is in
excellent agreement with the experimental value of 0.062.

\section{Conclusions}

We have derived a simple, compact expression for the Coulomb
contribution to the thermodynamic pressure in a system treated with
Ewald lattice summation.  For a system of point ions, we recover the
well-known relation between the pressure and potential energy.  We
then derive an expression for the pressure in a system of rigid
molecules carrying point charges.  The pressure in such a system can
be calculated from the total energy and the forces at each site alone.
This makes the implementation of that pressure formula trivial,
because both energy and forces are produced routinely in molecular
dynamics codes.  Moreover, these formulas are entirely independent of
the particular method used to resum the Coulomb interactions.  Ewald
summation, particle-mesh Ewald,\cite{Darden:93} kubic-harmonic
expansions,\cite{Slattery:80} and Lekner
sums\cite{Lekner:91,Jensen:96,Jensen:97:2} can be used readily.  For
approximate reaction-field
methods,\cite{Barker:73,Hummer:92:b,Hummer:94:e} expressions for the
pressure are suggested by analogy.

We have compared the thermodynamic pressure, obtained from the volume
dependence of the Helmholtz free energy, with the mechanistic virial
pressure.  We find that for rigid SPC/E water at standard conditions, the two
pressures are approximately equal (within errors of about 2 MPa) for
systems larger than $N=256$ water molecules.  For smaller systems, the
virial pressure exhibits a pronounced system-size dependence, whereas
the thermodynamic pressure can be calculated accurately by using as
few as 64 SPC/E water molecules.

\acknowledgments

Portions of this work were done under the auspices of the U.S.
Department of Energy.  This work was supported through a Los Alamos
National Laboratory LDRD grant.

\begin{figure}[htbp]
  \caption{Pressure of SPC/E water as a function of the inverse number
    of water molecules, $1/N$.  Cross symbols and dashed lines
    correspond to the virial pressure $p_V$.  Plus symbols and solid
    lines show the results for the thermodynamic pressure $p_T$.  The
    inset highlights results for larger system sizes, $N\geq 64$.
    Error bars indicate one standard deviation of the mean, estimated
    from a block error analysis.}
  \label{fig:p}
\end{figure}

\begin{figure}[htbp]
  \caption{Radial distribution functions of water atoms.  Oxygen-oxygen
    (top panel), oxygen-hydrogen (middle panel), and hydrogen-hydrogen
    (bottom panel) radial distribution functions are shown for
    different numbers of water molecules.  Arrows indicate half the
    box length, $r=L/2$, for different system sizes.  The radial
    distribution functions were calculated for distances
    beyond $L/2$ using appropriate weights.}
  \label{fig:gXX}
\end{figure}

\begin{table}[htbp]
  \caption{Characteristics and results of the Monte Carlo simulations
    of SPC/E water.  Statistical errors are one estimated
    standard deviation of the mean.  Also included are the cross-correlation
    coefficient $r$ and the absolute deviation $\Delta$, as defined
    in Eq.~(\protect\ref{eq:deviation}).}
  \begin{tabular}{rrrrrlr}
    $N$ & passes [$10^3$] & $\langle U/N \rangle$ [kJ/mol] & $p_V$
    [MPa] & $p_T$ [MPa]  & $r$ & $ \Delta$ [MPa]\\ \hline
    16  & 500  & $-46.95\pm 0.10$ & $  1061\pm 15 $&$   -7\pm 17  $ & 0.90 & 114\\
    32  & 900  & $-46.67\pm 0.04$ & $ 274.9\pm 6.0$&$-28.6\pm 5.6 $ & 0.95 &  51\\
    64  & 2100 & $-46.82\pm 0.03$ & $  52.9\pm 2.4$&$ -4.1\pm 2.4 $ & 0.98 &  24\\
    128 & 3000 & $-46.83\pm 0.02$ & $   2.8\pm 1.6$&$ -5.9\pm 1.6 $ & 0.990&  12\\
    256 & 1200 & $-46.79\pm 0.02$ & $  -3.8\pm 1.5$&$ -5.5\pm 1.5 $ & 0.995&   6\\
    512 & 540  & $-46.82\pm 0.02$ & $  -4.1\pm 1.7$&$ -4.4\pm 1.7 $ & 0.997&   3\\
  \end{tabular}
  \label{tab:p}
\end{table}
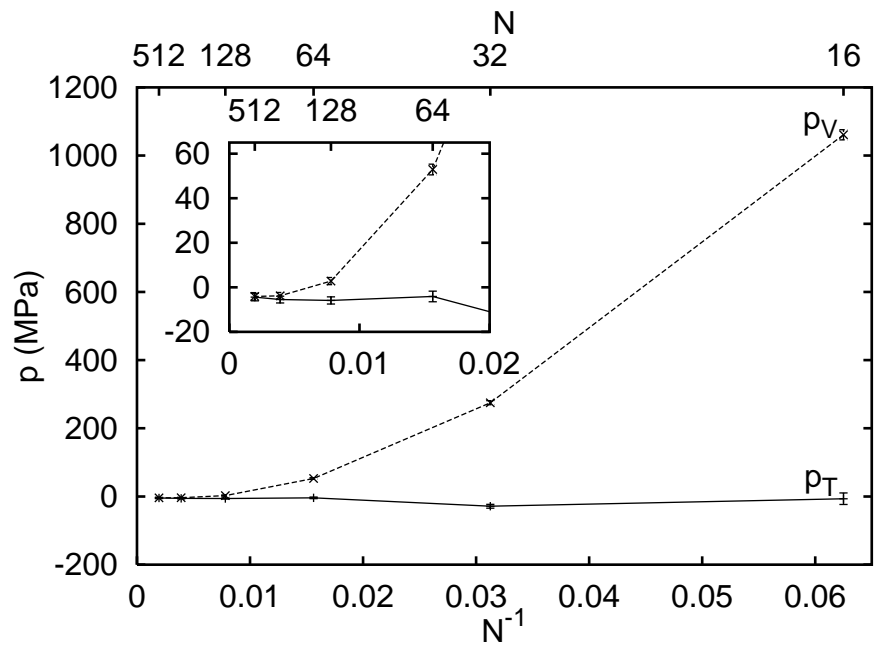
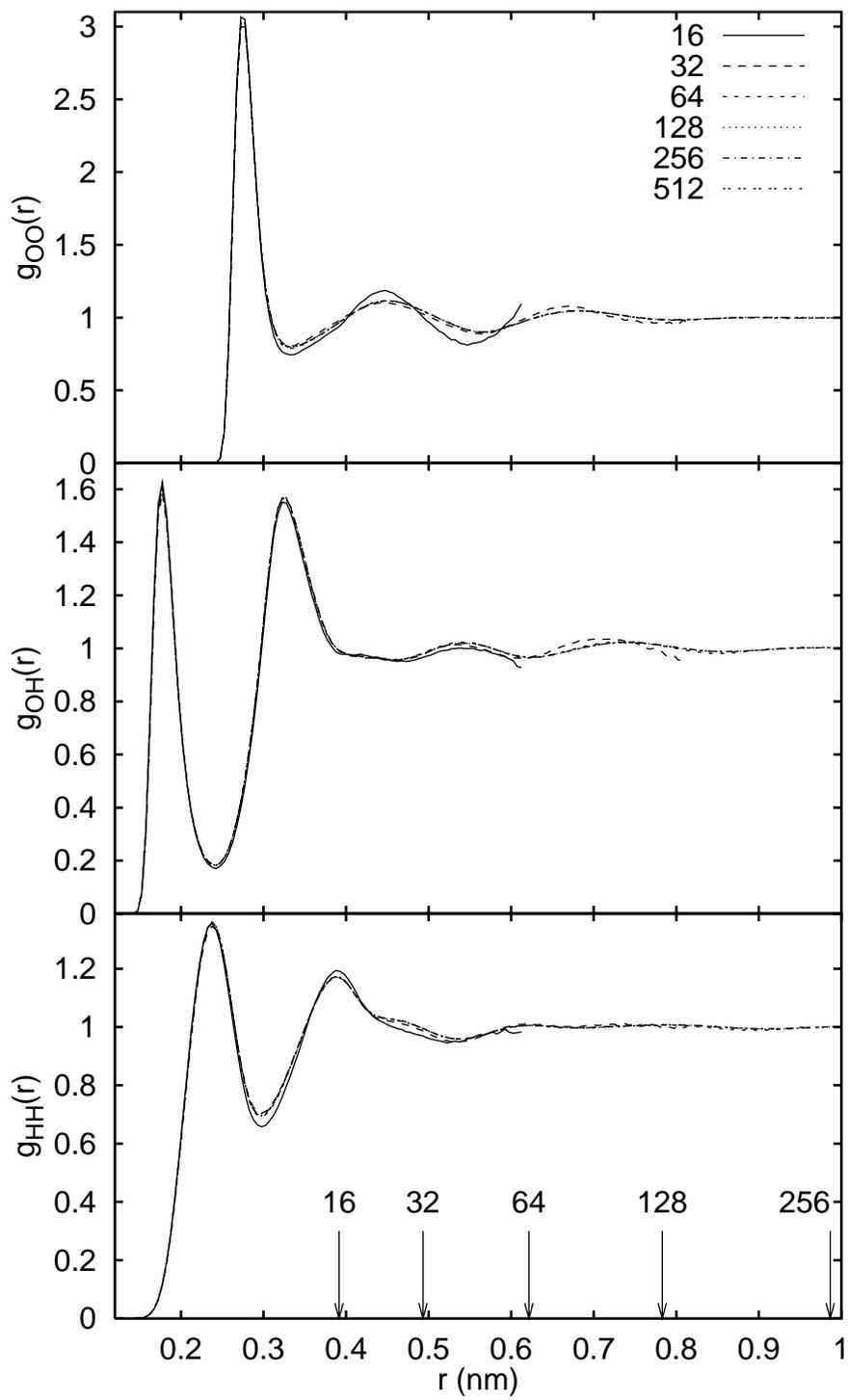
\end{document}